\documentclass[9pt, conference]{IEEEtran}
\IEEEoverridecommandlockouts
\usepackage{cite}
\usepackage{amsthm}
\usepackage{amsmath,amssymb,amsfonts}
\usepackage{tabularx}
\usepackage{fancyhdr}
\usepackage{tikz}
\usetikzlibrary{shapes.geometric}
\usepackage[normalem]{ulem}
\usepackage{url}
\usepackage{tikz}
\usepackage{tablefootnote}
\usepackage[bottom=2.1cm, top=1.20cm,left=1.72cm,right=1.72cm]{geometry}
\usepackage{amsthm}
\usepackage{amsmath,amsfonts}
\usepackage{algorithmic}
\usepackage{graphicx}
\usepackage{textcomp}
\usepackage[table,xcdraw]{}
\usepackage{booktabs} 
\usepackage{stfloats}
\usepackage[normalem]{ulem}
\usepackage{subfigure}
\usepackage{multirow}
\usepackage{paralist}
\usepackage{balance}
\usepackage{bm}
\usepackage[font=small]{caption}
\usepackage{newtxtext,newtxmath}
\usepackage{float}
\usepackage{booktabs}
\usepackage[linesnumbered, ruled, vlined]{algorithm2e}
\usepackage{xcolor}
\usepackage{amsmath} 
\usepackage{setspace}
\usepackage{caption}
\usepackage{subcaption}

\def\BibTeX{{\rm B\kern-.05em{\sc i\kern-.025em b}\kern-.08em
    T\kern-.1667em\lower.7ex\hbox{E}\kern-.125emX}}

\newcommand{\eg}{e.g.\xspace}

\newcommand{\defn}[1]{\textbf{\textit{#1}}}

\newrobustcmd*\blacka[1]{\tikz[baseline=(char.base)]{
            \node[shape=circle,draw,inner sep=1pt,fill,text=white,minimum size=1em] (char) {\textsf{\small a}};}}

\newrobustcmd*\blackb[1]{\tikz[baseline=(char.base)]{
            \node[shape=circle,draw,inner sep=1pt,fill,text=white,minimum size=1em] (char) {\textsf{\small b}};}}

\newrobustcmd*\blackc[1]{\tikz[baseline=(char.base)]{
            \node[shape=circle,draw,inner sep=1pt,fill,text=white,minimum size=1em] (char) {\textsf{\small c}};}}

\newrobustcmd*\blackd[1]{\tikz[baseline=(char.base)]{
            \node[shape=circle,draw,inner sep=1pt,fill,text=white,minimum size=1em] (char) {\textsf{\small d}};}}

\newrobustcmd*\blacke[1]{\tikz[baseline=(char.base)]{
            \node[shape=circle,draw,inner sep=1pt,fill,text=white,minimum size=1em] (char) {\textsf{\small e}};}}

\newrobustcmd*\blackf[1]{\tikz[baseline=(char.base)]{
            \node[shape=circle,draw,inner sep=1pt,fill,text=white,minimum size=1em] (char) {\textsf{\small f}};}}

\newrobustcmd*\blackg[1]{\tikz[baseline=(char.base)]{
            \node[shape=circle,draw,inner sep=1pt,fill,text=white,minimum size=1em] (char) {\textsf{\small g}};}}

\SetNlSty{textbf}{(}{)}
\begin{document}

\title{FlexStep: Enabling Flexible Error Detection in Multi/Many-core Real-time Systems}

\author{
    Tinglue Wang\IEEEauthorrefmark{2}, 
    Yiming Li\IEEEauthorrefmark{2}, 
    Wei Tang\IEEEauthorrefmark{2}, 
    Jiapeng Guan\IEEEauthorrefmark{3}, 
    Zhenghui Guo\IEEEauthorrefmark{2}, 
    Renshuang Jiang\IEEEauthorrefmark{4}, \\
    Ran Wei\IEEEauthorrefmark{5}\IEEEauthorrefmark{1},
    Jing Li\IEEEauthorrefmark{6}\IEEEauthorrefmark{1}, 
    Zhe Jiang\IEEEauthorrefmark{2}\IEEEauthorrefmark{1}\\
    \IEEEauthorblockA{\IEEEauthorrefmark{2}Southeast University, China.
    \IEEEauthorrefmark{3}Dalian University of Technology, China.
    \IEEEauthorrefmark{4}National University of Defense Technology, China.\\
    \IEEEauthorrefmark{5}Lancaster University, UK.
    \IEEEauthorrefmark{6}New Jersey Institute of Technology, US.
    }
    \thanks{\IEEEauthorrefmark{1} represents corresponding authors. Emails: r.wei5@lancaster.ac.uk, 	jingli@njit.edu and zhejiang.uk@gmail.com.}
}

\maketitle

\begin{abstract}
Reliability and real-time responsiveness in safety-critical systems have traditionally been achieved using error detection mechanisms, such as LockStep, which require pre-configured checker cores, strict synchronisation between main and checker cores, static error detection regions, or limited preemption capabilities. 
However, these core-bound hardware mechanisms often lead to significant resource over-provisioning, and diminished real-time responsiveness, particularly in modern systems where tasks with varying reliability requirements are consolidated on shared processors to improve efficiency, reduce costs, and save power. 
To address these challenges, this work presents FlexStep, a systematic solution that integrates hardware and software across the SoC, ISA, and OS scheduling layers. 
FlexStep features a novel microarchitecture that supports dynamic core configuration and asynchronous, preemptive error detection. 
The FlexStep architecture naturally allows for flexible task scheduling and error detection, enabling new scheduling algorithms that enhance both resource efficiency and real-time schedulability.
We publicly release FlexStep's source code, at \textbf{\url{https://anonymous.4open.science/r/FlexStep-DAC25-7B0C}}.
\end{abstract}

\section{Introduction}
\label{intro}

In safety-critical systems, such as automotive and avionics applications, processor cores are the foundational components ensuring reliable system operations.
These cores must be capable of detecting and correcting faults during task execution~\cite{rausand2014reliability,kumar2018reliability,maurya2020reliability,singh2021reliability,kumar2022reliability}, while also guaranteeing task schedulability~\cite{huang2014scheduling,ali2019rt-scheduling,kumar2020dynamic-scheduling,melani2017static-scheduling,calvaresi2018local-scheduling}.
However, since reliability and schedulability address distinct safety dimensions, they are usually achieved through different phases of the processor and system development and across various architectural levels.
Reliability is often ensured through hardware-level error detection mechanisms, such as the LockStep technology used in ARM Cortex R series processors~\cite{dual-lock,arm-triple-lock,iturbe2019arm,werdmuller2015addressing,iturbe2018addressing-lock,Cortex-R5}. While software fault tolerance mechanisms exist~\cite{hukerikar2018redthreads-soft,bernardi2015development-soft,sofycomputation-soft,thomas2013error-soft}, they offer limited coverage and lead to significant performance degradation.
Conversely, schedulability is ensured through scheduling algorithms implemented in the Operating System (OS)~\cite{ismael2021scheduling-SA,agrawal2021cpu-SA, hamayun2015optimized-SA}. 

\begin{figure}[t]
\begin{center}
\subfigure[LockStep configuration with fixed main core 0 \& checker core 1.]{
\includegraphics[width=\linewidth]{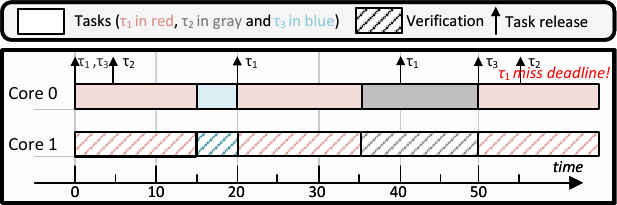}
\label{fig1:lockstep}
}
\subfigure[HMR~\cite{rogenmoser2023hybrid} configuration with limited flexibility and synchronous checking.]{
\includegraphics[width=\linewidth]{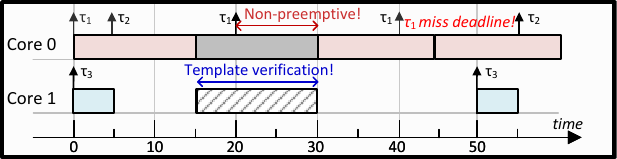}
\label{fig1:HMR}
}
\subfigure[FlexStep allowing asynchronous, selective, and preemptive checking.]{
\includegraphics[width=\linewidth]{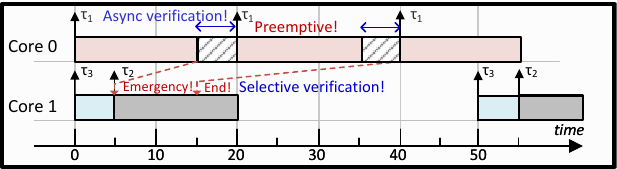}
\label{fig1:ours}
}
\caption{Scheduling on different dual-core architectures. Tasks $\tau_1, \tau_2, \tau_3$ have implicit deadlines and worst-case execution time (WCET) of 15, 15, 5, respectively. $\tau_1$ and $\tau_3$ are non-verification tasks that do not require error checking. An emergency occurs upon the arrival of the first job of task \( \tau_{2} \), requiring its first 10 units of work checked for errors.}
\label{scalability}
\end{center}
\end{figure}

\noindent \textbf{Existing works.}
As hardware computational capabilities advance and the diversity of software applications expands, there is a growing demand to execute multiple tasks with varying reliability requirements on shared processor cores\cite{6899155-trend,el2013across-trend2,saidi2015shift-trend3,cerrolaza2020multi-trend4,perez2022gpu-trend5,durrieu2014predictable-trend6}.
In this context, traditional LockStep technology faces significant limitations due to its rigid hardware design.
LockStep statically binds one or more identical cores, synchronously executing the same program and comparing the output results at each clock cycle.
This design mandates that all tasks running on LockStep cores undergo the highest level of error detection, irrespective of their actual reliability needs.
Consequently, this approach leads to inefficient use of error detection capabilities, suboptimal resource utilisation, significant power and area overheads, and challenges to system schedulability.
As illustrated in Fig.~\ref{fig1:lockstep}, under the rigid error detection mechanism of LockStep, even non-verification tasks like $\tau_1$ and $\tau_3$ are still subjected to error checking, consuming computational resources and power unnecessarily. This ultimately causes the third job of $\tau_1$ to miss its deadline.


Additionally, methods supporting reconfiguration, such as LockStep with split-lock~\cite{rogenmoser2023hybrid,Cortex-A76AE,kempf2022holistic-splitlock,kempf2021adaptive-flexible}, have been developed to reduce hardware resource overhead caused by rigid error detection.
For instance, the most recent Hybrid Modular Redundancy (HMR) approach~\cite{rogenmoser2023hybrid} explicitly separates mission-critical and performance-critical code sections. It leverages hardware architectural designs to enable runtime reconfiguration for split-lock, providing a degree of flexibility.
However, the inherent limitations of the ``core binding" design still impose significant constraints. Specifically, when reconfigured into verification mode, checker cores must perform error detection synchronously with the main core. Moreover, this synchronous error-checking execution cannot be preempted by non-verification tasks, even if those tasks have higher priorities or earlier deadlines. Additionally, HMR can only statically perform error checking for pre-determined tasks, lacking the capability to provide selective error checking based on dynamic system requirements. These constraints significantly reduce flexibility and negatively impact schedulability.
As shown in Fig.~\ref{fig1:HMR}, although HMR with runtime split-lock reduces resource wastage by not performing error checking for non-verification tasks $\tau_1$ and $\tau_3$ ($\tau_3$ can be executed on core 1), the ``core binding" design still prevents $\tau_1$ from preempting $\tau_2$'s error checking, resulting in $\tau_1$ missing its second deadline.

\begin{figure*}[t]
\begin{center}
\includegraphics[width=1\linewidth]{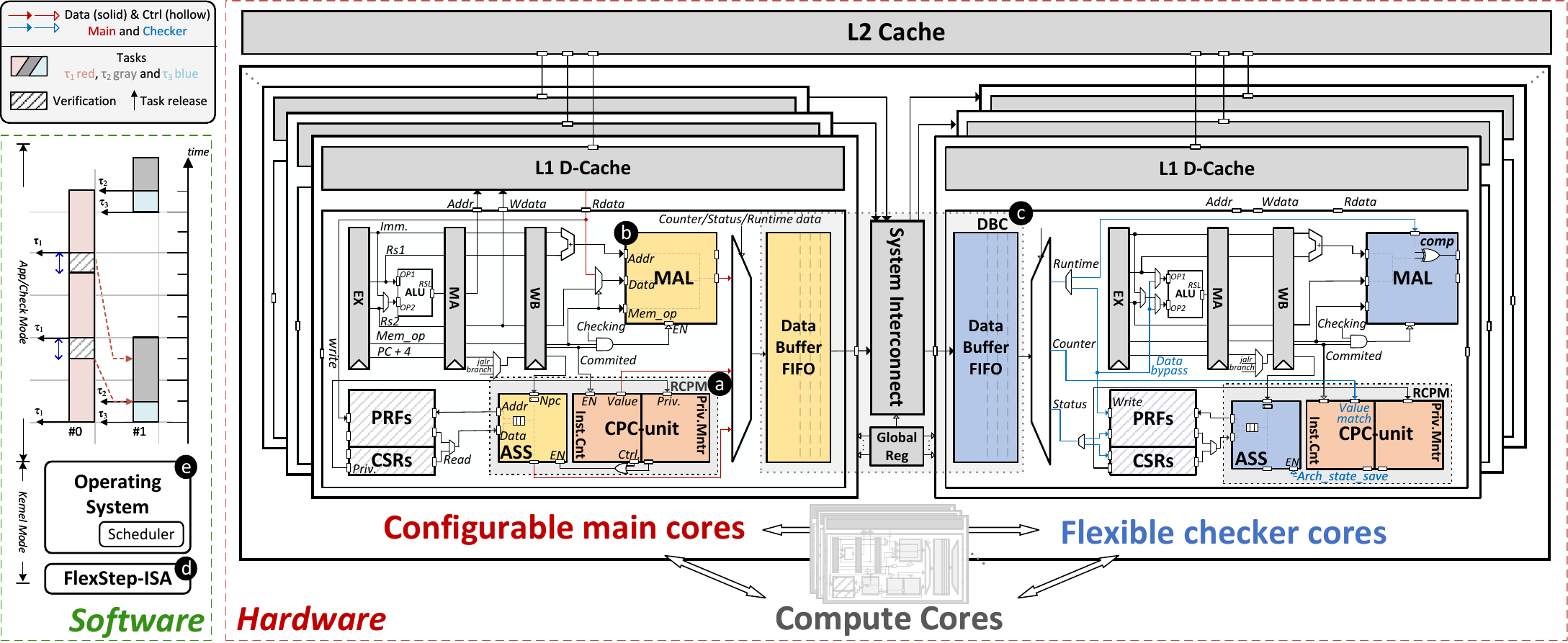}
\caption{FlexStep overview. At the hardware level (shown in red box), colored modules represent functional units added by FlexStep: orange hues denote identical functionality, while yellow and blue hues highlight variations in functionality for reconfigured main and checker cores, respectively. \blacka{a} Register Checkpoint Management (RCPM) manages Register Checkpoints and instruction counts, providing checker cores with execution boundaries and snapshots for verification (Sec.~\ref{RCPM}). \blackb{b} Memory Access Log (MAL) tracks and records memory accesses for correctness checks (Sec.~\ref{MAL}). \blackc{c} Data Buffering and Channelling (DBC) manages asynchronous data buffering and communication between cores via system interconnect (Sec.~\ref{DBC}). At the software level (shown in green box), FlexStep provides \blackd{d} customised ISA (Tab.~\ref{table:ISA}) to support the control interface between hardware microarchitecture and OS, enabling \blacke{e} a control flow capable of performing context switching between verification and non-verification tasks and more flexible scheduling.}
\label{Fig.overall}
\end{center}
\end{figure*}


\noindent \textbf{Contributions.}
To provide reliability assurance and schedulability guarantees for safety-critical systems while maintaining resource efficiency, the development of a more flexible hardware error detection architecture has become an industry necessity~\cite{jeffery2010flexible,bas2021safede-flexible, kempf2021adaptive-flexible,barbirotta2024dynamic-flexible}.
As illustrated in Fig.~\ref{fig1:ours}, such an architecture should decouple cores from the rigid LockStep configuration and employ a flexible error detection mechanism. This mechanism, with OS support, should enable \textit{error detection that is asynchronous, selective, and preemptable} by non-verification tasks. 
However, achieving this level of flexibility and efficiency presents significant technical challenges. 

To address these challenges, we present FlexStep, a comprehensive full-stack framework that integrates both hardware and software to deliver high flexibility and configurability. It implements a novel configurable microarchitecture supporting asynchronous, thread-level error detection between any cores. This approach overcomes the limitations of synchronous, core-binding methods, enabling more flexible error detection. Additional OS-layer controls are introduced to manage tasks with varying reliability requirements, ensuring seamless and efficient task scheduling.
Specifically, our contributions are:
\begin{itemize}
\item We modified the microarchitecture based on an open-source in-order core, Rocket~\cite{asanovic2016rocket}, to enable dynamic configuration and asynchronous, thread-level error detection mechanism. This lays the foundation for flexible and efficient scheduling at the OS layer.

\item We abstracted the hardware microarchitecture and configuration methods into a control flow featuring a customised Instruction Set Architecture (ISA), which is integrated into the context switch of OS scheduling. This enables the OS to reconfigure the hardware for task switching and preemption at runtime.
\item We proposed a full-stack framework that incorporates our multicore architecture and customised ISA for the OS scheduler, forming a comprehensive solution for safety-critical systems.

\item We formalised a new theoretical model for flexible error detection and developed scheduling algorithms for FlexStep to exploit its flexibility and improve real-time schedulability.
\end{itemize}

We implemented FlexStep on an AMD Alveo U280 FPGA and evaluated it using various metrics. Results demonstrate that FlexStep achieves microsecond-level detection latency with a 1.07\% slowdown, 2.21\% area overhead, and 2.89\% power consumption while delivering significantly improved schedulability compared to LockStep and HMR. 

\section{FlexStep: A Hardware/Software Co-design Framework}
\label{se: overall}



In FlexStep, any processor core can be configured as either a main core or a checker core. 
The main core executes software programs like a standard application core, while the checker core verifies the correctness of the main core through a second run of the same program.
Fig.~\ref{Fig.overall} illustrates the FlexStep framework. 

Unlike the synchronised verification in LockStep, FlexStep employs error detection based on {Register Checkpoints} ({RCPs}) -- architectural states captured at specific points -- and memory accesses, similar to Paramedic~\cite{ainsworth2018parallel,ainsworth2021paradox,ainsworth2019paramedic}. 
Specifically, a user thread running on the main core is divided into small \defn{checking segments}, which can be reproduced on its associated checker core(s). Each segment is identified by \defn{Start Register Checkpoints} (\defn{SCPs}) and \defn{End Register Checkpoints} (\defn{ECPs}) stored in the Register Checkpoint Management units (Fig.~\ref{Fig.overall}.\blacka{a}, Sec.~\ref{RCPM}).
The checker core halts memory access and sequentially replays the checking segments. Before executing a checking segment, it initialises its architectural state to the segment's SCP and compares its final state to the ECP after execution. During execution, memory access data (e.g., addresses and data for \texttt{LD/ST}, \texttt{LR/SC}, \texttt{AMO} instructions) recorded in the Memory Access Log (Fig.~\ref{Fig.overall}.\blackb{b}, Sec.~\ref{MAL}) are used for replay execution and verified at runtime. The program execution on the main core is deemed correct if the ECPs of all checking segments and memory access data match the original execution.

The fundamental idea is that, as long as all data related to Register Checkpoints and memory accesses performed by the thread -- required for reproducing execution and verification -- are recorded and buffered temporarily, the checker thread can be executed \textit{asynchronously} on any other core in the system.
Therefore, a checker core can execute other tasks instead of immediately starting verification, as long as the necessary data of the verification task have been extracted and buffered from the main core. This design improves both utilisation and flexibility. 

FlexStep allows user threads running on any core to be duplicated and verified on different core(s) using a data buffer and a \defn{configurable interconnected channel} (Fig.~\ref{Fig.overall}.\blackc{c}, Sec.~\ref{DBC}). Unlike LockStep and HMR, which supports specific verification modes such as Dual-Core LockStep (DCLS) and Triple-Core LockStep (TCLS) by binding cores via a shared input path for instructions, data, and control, FlexStep’s interconnected channel can be configured to operate in one-to-one (similar to DCLS), one-to-two (similar to TCLS), or more modes. This accommodates task scenarios with varying safety-criticality demands.

To manage cores and schedule error detection within the system, FlexStep abstracts the underlying configurable hardware control interfaces as a \defn{customised ISA} (Fig.~\ref{Fig.overall}.\blackd{d}, Tab.~\ref{table:ISA}) and introduces a control flow integrated into the OS layer. 
The ISA defines the instructions specific to the main core or checker core and global instructions for both.
This design makes all \defn{core attributes} (main, checker, or compute) visible to the OS, allowing them to be dynamically configured at runtime to facilitate OS scheduling (Fig.~\ref{Fig.overall}.\blacke{e}). 
Preemptive execution is supported, allowing any core to be interrupted at any time and enabling more urgent tasks to be executed and meet their deadlines. 
Thus, the OS can dynamically configure core attributes and verification modes as needed, leveraging asynchronous verification and task preemption to optimise real-time scheduling while ensuring system reliability.

\section{Microarchitecture}

To implement FlexStep at the hardware level, modifications to the microarchitecture are required, incorporating several key functional units, as illustrated in Fig.~\ref{Fig.overall}. 
For demonstration, we used the Rocket core as a case study to implement our framework and evaluate its effectiveness. It is important to note that incorporating the same functional units into each core is essential to enable dynamic switching between main and checker attributes. However, this requirement poses constraints on processor area availability for extensions. To address this, cores with different attributes must maximise the reuse of functional units to improve utilisation and minimise area costs. This section discusses the technical details of the modified microarchitecture.






\begin{table}[t]
\centering
\caption{FlexStep ISA, abstracting control interface for software.}
\label{table:ISA}
\resizebox{0.935\columnwidth}{!}{%
\begin{tabular}{l|l}
\bottomrule
\hline
\textbf{Instruction}&  \multicolumn{1}{c}{\textbf{Description}}\\
\hline
\texttt{\textbf{G.IDs.contain}} & Return core attributes (Main/Checker)\\
\hline
\texttt{\textbf{G.Configure}} & Configure the main and checker cores'ID\\
\hline
\texttt{\textbf{M.associate}} & Allocate one or multiple checker core(s) to main\\
\hline
\texttt{\textbf{M.check}} & Enable/Disable the checking function\\
\hline
\texttt{\textbf{C.check\_state}}& Switch the checking state (busy/idle)\\
\hline
\texttt{\textbf{C.record}}& Record the context to ASS\\
\hline
\texttt{\textbf{C.apply}}& Apply the SCP from data channel\\
\hline
\texttt{\textbf{C.jal}}& Jump to the next pc (npc) of SCP\\
\hline
\texttt{\textbf{C.result}}& Return the comparison result\\
\hline
\toprule
\end{tabular}
}
\end{table}

\subsection{Register Checkpoint Management Units (RCPM)}\label{RCPM}
RCPM (Fig.~\ref{Fig.overall}.\blacka{a}) consists of the \defn{Checkpoint Control} (\defn{CPC}) and \defn{Architectural State Snapshot} (\defn{ASS}) units. CPC controls the start and end of a checking segment. In a checker core, CPC identifies segment boundaries and checks the correctness of the execution. ASS is a storage unit, which captures Register Checkpoint snapshots and releases them for either transmitting or applying under the control of CPC.

\newcommand*\numcircledtikz[1]{%
    \tikz[baseline=(char.base)]{%
        \node[shape=circle,draw,inner sep=0.5pt,minimum size=1em] (char) {\normalsize #1};%
    }%
}

\begin{figure}[t]
    \begin{center}
    \includegraphics[width=0.9\linewidth]{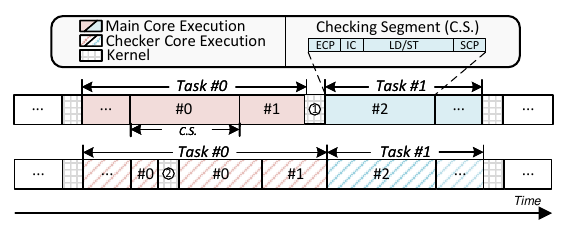}
    \caption{Checking segments: their components, main and checker execution of them, and them of default length or interrupted by kernel.
    }
    \label{fig3:a}
    \end{center}
\end{figure}

Checkpoint Control (CPC) includes an instruction counter (Fig.~\ref{Fig.overall}. Inst.Cnt) and a privilege-level mode monitor (Fig.~\ref{Fig.overall}. Priv.Mntr). The FlexStep architecture only supports user mode checking, and all cores could enter kernel mode during execution, resulting in premature extermination (Fig.~\ref{fig3:a}.\numcircledtikz{1}) and temporary deviation (Fig.~\ref{fig3:a}.\numcircledtikz{2}) of a checking segment. Thus, it is crucial to accurately count the number of instructions executed in user mode by both the main and checker cores to ensure that they execute the same length of a checking segment.  

\noindent \textbf{CPC's Working Mechanism.} For the main core, a new checkpoint is generated under the following conditions: a) a privilege level mode switch occurs; 
b) an instruction count limit is reached (default is 5000). At the ECP, the counter stops and records the number of instructions of the entire segment. Throughout the segment, the main core sends SCP, \texttt{LD/ST} log, instruction count (IC), and ECP in order (Fig.~\ref{fig3:a}) for the checker core to receive.
For the checker core, upon applying the SCP, it begins checking, until the instruction count reaches the same value as main core. It then uses its own architectural state to verify the ECP.

Architectural State Snapshot (ASS) is responsible for temporarily storing Register Checkpoints and general architectural states to facilitate duplicate execution and rapid execution state switching. Before the main core sends SCPs and ECPs, the ASS temporarily stores them and organises them into a format suitable for receiving by the checker core before forwarding them. In the checker core, before the checker thread enters actual duplicate execution, it stores the architectural state in ASS instead of main memory, allowing for rapid extraction of the state upon reaching an ECP and returning to fetch the pending SCP in time.  


\subsection{Memory Access Log Unit (MAL)}\label{MAL}


For the main core, information required for both duplicate execution and verification needs to be recorded and transmitted in order during checking segments. MAL (Fig.~\ref{Fig.overall}.\blackb{b}) identifies a memory instruction, records its data and addresses in the decode stage, and pipelines them to the commit stage, where it will be packaged if the commit is valid. It ensures that the package follows the order of instructions committed when entering the data channel. As for the checker core, MAL monitors memory access during the duplicate execution, and records the information that needs to be verified in the same way. When instructions are committed, the recorded information is compared against that retrieved from the main core, and the result is reported.

\noindent \textbf{Handling memory instructions with multiple micro-ops.} For regular \texttt{LD/ST} instructions, the log unit packages them into a single entry. However, for instructions with multiple memory micro-operations (\texttt{LR}, \texttt{SC} and \texttt{AMO}), the relevant data are packaged into multiple entries to minimise data width. This requires slight modifications to the processing of such instructions in the checker core and introduces additional latency. Nevertheless, the simplification of data format and reduction of storage overhead outweigh the disadvantages, as these instructions account for only a small fraction in most workloads.

\subsection{Data Buffering and Channelling Units (DBC)}\label{DBC}



DBC (Fig.~\ref{Fig.overall}.\blackc{c}) consists of \defn{Data Buffer FIFO} to buffer data of checking segments for conflict resolution and asynchronous error checking and \defn{System Interconnect} that establishes and alters links of the FIFOs between main cores and their associated checker cores.

\noindent \textbf{Configurable Inter-core Channels.} Inter-core communication channels between main and checker cores are achieved through the System Interconnect, a fully connected MUX-DEMUX network laid out between FIFO of each core. A global register, which can be modified by custom ISA instructions, generates control signals for MUX/DEMUX to establish channels between the main core and one or more checker cores. This design eliminates communication conflicts and ensures minimal communication latency at a small scale. As the design scales up, the wiring complexity grows exponentially, and the interconnect could be replaced with a bus interconnect or NoC.

\noindent \textbf{Buffering for Conflict Resolving and Asynchrony.} In line with previous work~\cite{ainsworth2018parallel,ainsworth2019paramedic,ainsworth2021paradox}, we adopt an SRAM-based FIFO for data buffering, allowing the checker core to retrieve the data and addresses of memory instructions for duplicate execution directly from the FIFO rather than the main memory. This design provides buffering essential for the proposed asynchronous verification. The larger the FIFO capacity of storing \texttt{LD/ST} entries, the longer the checker thread can lag behind the associated main thread, thereby providing more scheduling flexibility. 
The main core’s FIFO is used to resolve conflicts when two main cores compete for access to a checker core. In such cases, only one main core's FIFO is permitted to send data to the checker core, while the other temporarily buffers its data in its own FIFO until the checker core is released. To further support conflict resolution and asynchronous operation, additional buffering can be allocated in main memory, accessed via DMA, to provide the extra space needed.

\setlength{\textfloatsep}{0pt}

\begin{algorithm}[t]
\small
\setstretch{0.95} 
\SetAlgoNlRelativeSize{0}
\SetNlSty{}{}{}
\SetAlgoNlRelativeSize{-1}
\DontPrintSemicolon
\SetAlgoLined
\SetKwProg{Fn}{Function}{:}{}
$\vartriangleright$ \textbf{\texttt{Scheduler}}\\
\SetKwFunction{FMain}{\normalfont Context\_Switch}
    \SetKwProg{Fn}{Function}{:}{}
    \Fn{\FMain{\normalfont task \textit{*current}, core \textit{core\_id}}}{
    {
        \uIf{\normalfont (\emph{\textcolor{blue}{G.Main\_IDs.contain(\textcolor{black}{core\_id})}})}{
            \textcolor{blue}{M.check.disable();}
        }
        \ElseIf{\normalfont (\emph{\textcolor{blue}{G.Checker\_IDs.contain(\textcolor{black}{core\_id})}})}{
            \textcolor{blue}{C.check\_state(idle);}
        }
        
        \textbf{Kernel}.Intr(DISABLE);\\
        task *\textit{next} = NULL;\\
        \emph{/* Switch to the next task */}\\
        \textbf{Kernel}.Context.save(\emph{current});\\
        \emph{next} = \textbf{Kernel}.Find\_next();\\
        
        \eIf{\normalfont (\emph{next$\rightarrow$new\_release})}{
            \emph{/* Configure the main and checker core's ID */}\\
            \textbf{\textcolor{blue}{G.Configure(Main\_IDs, Checker\_IDs);}}\\
            \textbf{Kernel}.Context.init(\emph{next});
        }
        {
            \textbf{Kernel}.Context.restore(\emph{next});
        }
        
        \emph{current} = \emph{next};\\
        \textbf{Kernel}.Intr(ENABLE);\\
        \uIf{\normalfont (\emph{\textcolor{blue}{G.Main\_IDs.contain(\textcolor{black}{core\_id})}})}{
            \emph{/* Associate checker cores and enable checking */}\\
            \textcolor{blue}{M.associate(Checker\_ID(s));}\\
            \textcolor{blue}{M.check.enable();}
        }
        \ElseIf{\normalfont (\emph{\textcolor{blue}{G.Checker\_IDs.contain(\textcolor{black}{core\_id}}}) \textbf{and} \normalfont \emph{next$\rightarrow$checker\_thread})}{
            \textcolor{blue}{C.check\_state(busy);}
        }
        
        \textbf{Kernel}.Context.jalr(\emph{current$\rightarrow$pc});
    }
}
\textbf{End Function} \\
\caption{Each core's context switch.}
\label{al:context_switch}
\end{algorithm}

\section{New ISA and OS Kernel Add-ons}
In the FlexStep framework, the OS plays a pivotal role in managing hardware resources and ensuring real-time schedulability. To enhance its functionality, we leverage a control interface within the ISA (Tab.~\ref{table:ISA}), enabling efficient core management with minimal modifications to the OS. This approach ensures compatibility with existing application programs while improving overall scheduling performance.

\subsection{New ISA Support}
As shown in Tab.~\ref{table:ISA}, the new ISA is classified into three categories for the main core (\texttt{M.x()}), checker core (\texttt{C.x()}) and both of them (\texttt{G.x()}). Global instructions include an \texttt{IDs.contain} that can identify the attributes of each core and a \texttt{Configure} which is used to write the IDs of main and checker cores into global configuration registers. For main cores, \texttt{associate} allocates one or multiple checker cores for redundant execution, and \texttt{check} controls the checking capability by managing the relevant units. For checker cores, we present a \texttt{check\_state} for changing the checking state of the core, and a \texttt{record} for recording the architectural state into the ASS. In addition, a pair of ``atomic" instructions \texttt{apply} and \texttt{jal} are presented to replay the checking segment on checker cores. The \texttt{apply} is used to update the architectural registers, while the \texttt{jal} is used to jump to the main thread, which is modified from the standard jump instruction with a designated target from the main cores for accurate branch misprediction handling without necessitating any changes to the microarchitecture. Finally, the \texttt{result} reports the comparison result at each ECP.

\subsection{OS Kernel Add-ons and Customised Checker Thread}

The modification of OS (Fig.~\ref{Fig.overall}.\blacke{e}) is shown in Al.~\ref{al:context_switch}, which only requires adding a few lines of code to the context switch function of the scheduler with the new ISA. When entering a context switch, each core executes different instructions based on its attributes to switch off the checking function (Al.~\ref{al:context_switch}: lines 3 - 7). Additionally, when a new task is released, the \texttt{Configure} is executed for (re-)configuring the global registers (Al.~\ref{al:context_switch}: line 15). Finally, for the main core, the \texttt{associate} equips it with the specific checker core(s), while the \texttt{check\_enable} is called to switch on the checking function (Al.~\ref{al:context_switch}: line 25). For the checker core, if the next task is a verification task, it will switch its checking state to busy (Al.~\ref{al:context_switch}: line 27) and enter a checker thread.

\setlength{\textfloatsep}{-2pt}
\begin{algorithm}[t]
\small
\setstretch{0.95} 
\SetAlgoNlRelativeSize{0}
\SetNlSty{}{}{}
\SetAlgoNlRelativeSize{-1}
\DontPrintSemicolon
\SetAlgoLined
\SetKwFunction{FMain}{\normalfont Checker\_Thread}
    \SetKwProg{Fn}{Function}{:}{}
    \Fn{\FMain{}}{
    {
        \emph{/* launching checker thread with P-Thread */}\\
        ... \\
        \normalfont C.record(\emph{ASS});  \emph{// return position after checking}\\
        \If {\normalfont (\emph{!C.rslt()})}{
             ReportErr();
         }

        \textbf{while} {\normalfont (\emph{C.NewSCP() != ready});} \\
        C.apply(C.NewSCP.data);\\
        C.jal(C.NewSCP.npc);
    }
}
\textbf{End Function}
\caption{Customised checker thread.}
\label{al:checker_thread}
\end{algorithm}


Besides the general modification of the context switch for all cores, we have also developed a dedicated thread for checker cores (Al.~\ref{al:checker_thread}). Similar to the regular context switch, 
the \texttt{record} is used to store current architectural registers data in ASS for restoration after the check is completed (Al.~\ref{al:checker_thread}: line 4). Next, the checker core will enter a loop and continuously request new SCP from the FIFO (Al.~\ref{al:checker_thread}: line 8). Once a new SCP arrives, the \texttt{apply} will deploy it to the architectural registers (Al.~\ref{al:checker_thread}: line 9), followed by the \texttt{jal} to control the directional jump of PC (Al.~\ref{al:checker_thread}: line 10), thereby ensuring that the checker core executes the correct sequence of instructions based on the updated state. Lastly, the verification result is returned by the \texttt{result} (Al.~\ref{al:checker_thread}: lines 5 - 7), and the correction mechanism will be triggered if an error is detected.


\begin{algorithm}[t]
\small
\setstretch{0.95}
\SetNlSty{}{}{}
\SetAlgoNlRelativeSize{-1}
\DontPrintSemicolon
\caption{Scheduling with asynchronous verification.}
\label{alg:scheduling}
\textbf{Input:} $\Gamma$: Task set $\tau_i \in \{\mathcal{T}^N, \mathcal{T}^{V2}, \mathcal{T}^{V3}\}$, $m$ available cores\\
\textbf{Output:} Deadline-compliant task partitioning\\
\textbf{for} each core  $k \in \{1, \dots, m\}$ \textbf{do }$  \Delta[k] \gets 0$;\\
\For{\normalfont{each} $\tau_i \in \{\mathcal{T}^{V3}, \mathcal{T}^{V2}\}$}{
    \textbf{if} $\tau_i \in \mathcal{T}^{V2}$ \textbf{then} $D'_i \gets D_i / 2$; \\
    \textbf{else if} $\tau_i \in \mathcal{T}^{V3}$ \textbf{then} $D'_i \gets (\sqrt{2} - 1) D_i$;\\
    $\delta^{\text{o}}_{i} \gets C_i / D'_i$; 
    $\delta^{\text{v}}_{i} \gets C_i / (D_i - D'_i)$;\\

    $k \gets \text{argmin}_{k \in \{1, \dots, m\}} \Delta [k]$; \\
    Assign $\tau_i^o$ to core $k$;
    $\Delta [k] \gets \Delta [k] + \delta^{\text{o}}_i$;\\
    
    $k' \gets \text{argmin}_{k' \in \{1, \dots, m\} \setminus \{k\}} \delta [k']$;\\
    Assign $\tau_i^{v}$ to core $k'$;
    $\Delta [k'] \gets \Delta [k'] + \delta^{\text{v}}_i$;\\
    
    \If{$\tau_i \in \mathcal{T}^{V3}$}{
        $k'' \gets \text{argmin}_{k'' \in \{1, \dots, m\} \setminus \{k, k'\}} \Delta [k'']$;\\
        Assign $\tau_i^{v'}$ to core $k''$;
        $\Delta [k''] \gets \Delta [k''] + \delta^{\text{v}}_i$;\\
    }
}
\For{\normalfont{each} $\tau_i \in \mathcal{T}^N$}{
    $\delta^{\text{o}}_i \gets C_i / D_i$;\\
    $k \gets \text{argmin}_{k \in \{1, \dots, m\}} \Delta [k]$; \\
    Assign $\tau_i$ to core $k$;
    $\Delta [k] \gets \Delta [k] + \delta^{\text{o}}_i$;\\
}
\For{\normalfont{each core} $k \in \{1, \dots, m\}$}{
    \textbf{if} $\Delta [k] > 1$ \textbf{then} \Return Fail;\\
}
\Return Success;

\end{algorithm}

\section{System Model and Scheduling Analysis}
\label{theory}
We now describe the formal model and analysis for the scheduling problem with verification ensuring reliability and real-time guarantees. 

\noindent \textbf{Model.}
We seek to schedule a task set consisting of $n$ sporadic tasks $\tau_1,\cdots \tau_n$ on $m$ cores. Each task $\tau_i$ is characterised by its worst-case execution time $C_i$, period $T_i$, and implicit deadline $D_i = T_i$. 
Additionally, tasks can be classified into three types $\mathcal{T}^N$, $\mathcal{T}^{V2}$ and $\mathcal{T}^{V3}$, depending on whether they may require additional error checking during runtime to satisfy the reliability constraint. A task in $\mathcal{T}^N$ is a non-verification task that only needs to execute its normal work once every period and meet its deadline. In comparison, a task in $\mathcal{T}^{V2}$ may require double-check, while a task in $\mathcal{T}^{V3}$ may require triple-check online. In the event of an emergency, the system dynamically triggers additional error checking for one or more jobs of specific verification tasks based on the nature of the emergency. Double-check (or triple-check) verification involves duplicating the computation under verification once (or twice) and reproducing it on one (or two) cores distinct from the core performing the original computation. 

If asynchronous verification is supported, the duplicated computations can be executed after the original computation but must still be completed before the job deadline to ensure timely error detection and maintain system reliability. As discussed earlier, FlexStep supports both asynchronous verification and selective error checking, allowing error checking to be dynamically triggered for a task and performed on specific portions of a job or task as needed. This flexibility opens new opportunities to improve resource efficiency and schedulability when using finer-grained modeling of verification and developing tailored schedulers and analyses. As an initial step in this direction, we limit our focus on asynchronous verification only, while assuming that all jobs of verification tasks require full error checking and that all tasks and associated error checking must meet their deadlines. 
Without additional information, such as the probability of dynamic verification or assumptions allowing low-criticality tasks to miss deadlines during system emergencies, this formulation represents the system's worst-case behaviour. Experiments demonstrate that FlexStep can significantly enhance schedulability, even under this restricted worst-case system model, by only leveraging asynchronous verification.

\noindent \textbf{Scheduling algorithm and analysis.}
For this scheduling problem, we propose a simple scheduling algorithm and corresponding schedulability analysis based on the partitioned Earliest Deadline First (partitioned EDF). The algorithm partitions all non-verification tasks, verification tasks, and their duplicated computations for double-check or triple-check verification on $m$ available cores. 

Since error-checking computations cannot begin before the original computation, our algorithm assigns a virtual deadline $D'_i$ to each verification task to reserve enough time for verification. This virtual deadline is used for scheduling the original computation on its assigned core using EDF, while the original deadline is used for scheduling the duplicated computations. For a double-check task $\tau_i \in \mathcal{T}^{V2}$, $D'_i = D_i/2$, and for a triple-check task $\tau_i \in \mathcal{T}^{V3}$, $D'_i = (\sqrt{2}-1)D_i$. The virtual deadline is chosen to minimise the total density of the original and duplicated computations, optimising system schedulability. 
The densities of the original and error-checking  computations for verification task $\tau_i$ are $\delta^o_i = C_i/D'_i$ and $\delta^v_i = C_i/(D_i-D'_i)$, respectively. Note that, in practice, duplicated computations can start as soon as the original computation begins execution. However, our partitioning algorithm and schedulability analysis consider the worst-case scenario -- the error-checking computation starts only after the virtual deadline.

Our partitioning algorithm (Al.~\ref{alg:scheduling}) first partitions verification tasks with descending utilisation and ensures that their original and error-checking computations are allocated to different cores (Al.~\ref{alg:scheduling}: lines 4 - 14). Their calculated densities $\delta^o_i, \delta^v_i$ are used to update the total density of allocated cores. Non-verification tasks are then partitioned with descending utilisations (Al.~\ref{alg:scheduling}: lines 15 - 18). As EDF is used to execute tasks allocated to each core, core $k$ is deemed schedulable if its total density ($\Delta[k]$) of assigned tasks does not exceed one. The entire task set is guaranteed to be schedulable if all tasks (and their error-checking computations) are successfully assigned to cores and all cores remain schedulable (Al.~\ref{alg:scheduling}: lines 19 - 20). Since our schedulability test is a sufficient test, when the test fails and hard real-time guarantees are not required, we can remove the virtual deadline and use the verification task's original deadline and utilisation for scheduling and partitioning.

\section{Evaluation}
\label{evaluation}

We built FlexStep featuring homogeneous SoC with multiple Rockets and implemented the microarchitecture upon an open-source platform Chipyard~\cite{chipyard} (v1.8.0). With TSMC 28nm PDKs~\cite{TSMC_28}, we synthesised the RTL using Synopsys (v2019.12). We deployed the RTL design on the AMD Alveo
U280 FPGA using FireSim~\cite{karandikar2018firesim} to simulate the settings in Tab~\ref{table:Setup} and boot Linux kernel (v5.7), executing full
SPECint 06~\cite{henning2006spec} and Parsec V3~\cite{bienia2008parsec} with simmedium dataset.


\begin{figure*}[t]
\begin{center}
\subfigure[Parsec (v3.0).]{
\includegraphics[trim=0cm 0.1cm 0cm 0.1cm, clip, width=0.44\linewidth]{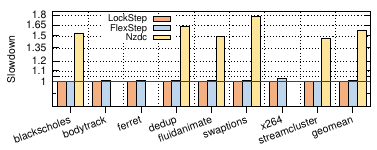}
\label{fig:perf_parsec}
}
\hspace{32pt}
\subfigure[Full SPECint CPU2006.
]{
\includegraphics[trim=0cm 0.1cm 0cm 0.1cm, clip, width=0.44\linewidth]{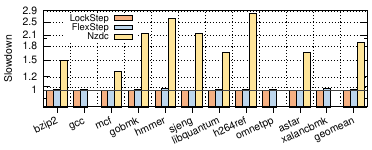}
\label{fig:perf_spec}
}
\caption{Performance slowdown of Parsec and SPEC06 using LockStep, FlexStep and Nzdc.}
\label{perf}
\end{center}
\end{figure*}

\begin{figure*}[t]
\begin{center}
\subfigure[$m=8, n=160, \alpha=6.25\%, \beta=6.25\%$.]{
\includegraphics[trim=0cm 0.2cm 0cm 0.2cm, clip, width=0.32\linewidth]{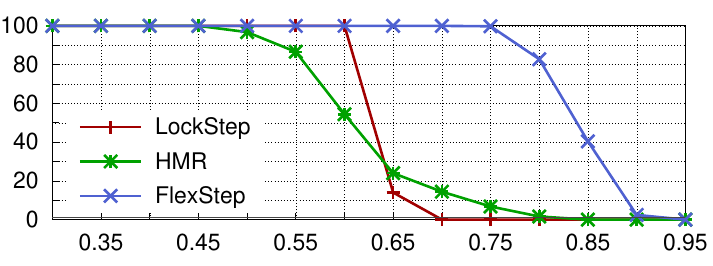}
\label{fig:160-8-10-10}
}
\subfigure[$m=8, n=160, \alpha=12.5\%, \beta=12.5\%$.]{
\includegraphics[trim=0cm 0.2cm 0cm 0.2cm, clip, width=0.32\linewidth]{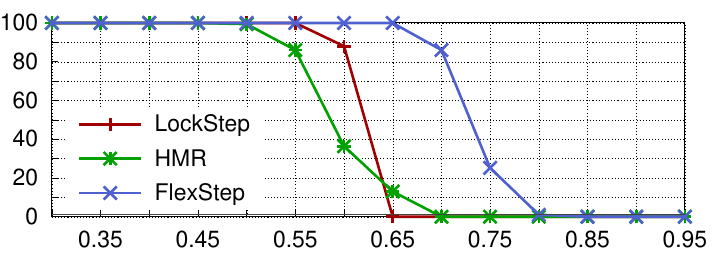}
\label{fig:160-8-20-20}
}
\subfigure[$m=8, n=160, \alpha=25\%, \beta=25\%$.]{
\includegraphics[trim=0cm 0.2cm 0cm 0.2cm, clip, width=0.32\linewidth]{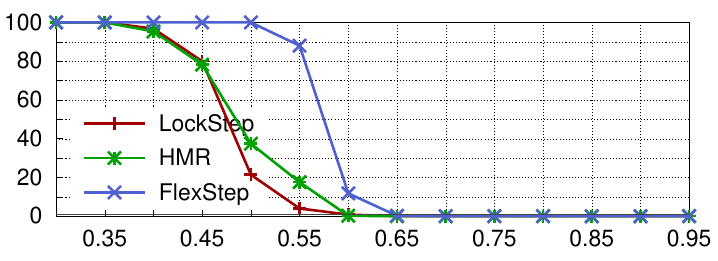}
\label{fig:160-8-40-40}
}
\subfigure[$m=8, n=160, \alpha=25\%, \beta=0\%$.]{
\includegraphics[trim=0cm 0.2cm 0cm 0.2cm, clip, width=0.32\linewidth]{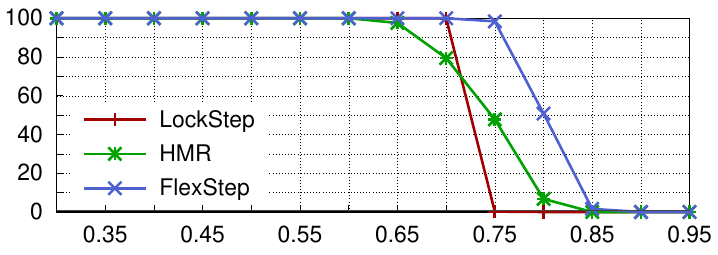}
\label{fig:160-8-0-40}
}
\subfigure[$m=16, n=160, \alpha=12.5\%, \beta=12.5\%.$]{
\includegraphics[trim=0cm 0.2cm 0cm 0.2cm, clip, width=0.32\linewidth]{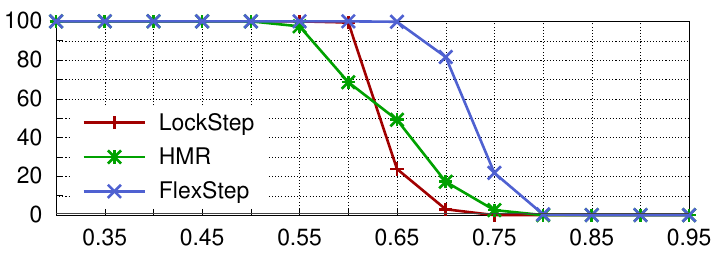}
\label{fig:160-16-20-20}
}
\subfigure[$m=8, n=80, \alpha=25\%, \beta=25\%.$]{
\includegraphics[trim=0cm 0.2cm 0cm 0.2cm, clip, width=0.32\linewidth]{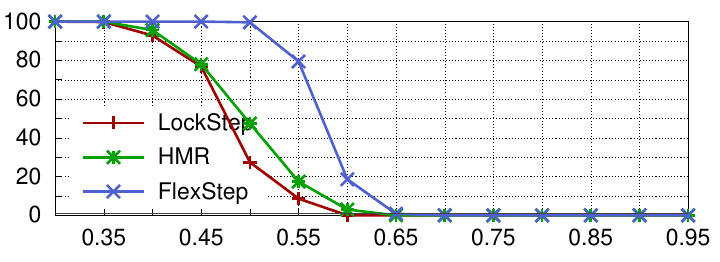}
\label{fig:80-8-20-20}
}
\caption{Percentage of schedulable task sets ($y$-axis) under LockStep, HMR, and FlexStep with increasing task set utilisations ($x$-axis) and varying system configurations: $m$ (number of cores), $n$ (number of tasks), 
$\alpha$ (percentage of double-check tasks), and $\beta$ (percentage of triple-check tasks).}
\label{sr}
\end{center}
\end{figure*}

\begin{figure}[t]
\center
\includegraphics[width=0.90\linewidth]{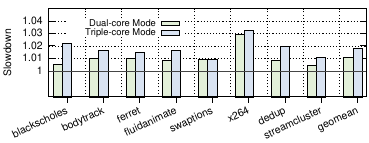}
\caption{The performance slowdown in dual-core and triple-core mode.}
\label{diffmode}
\end{figure}

\begin{table}[h]
\centering
\renewcommand{\arraystretch}{0.88}
\caption{Hardware configurations evaluated. 
}
\begin{tabular}{ll}
\multicolumn{2}{c}{\textit{Homogeneous Core}}     \\ \hline
Core         & \begin{tabular}[c]{@{}l@{}}In-order scalar Rocket, @1.6GHz\\ \end{tabular} \\
Pipeline     & \begin{tabular}[c]{@{}l@{}}5-stage pipeline, 64 Int/FP Phy Registers, \\1 ALU, 1 DIV, 1 FPU, 1 CSR \end{tabular}  \\ 
Branch Pred. & \begin{tabular}[c]{@{}l@{}}512-entry BHT, 28-entry BTB, 6-entry RAS\end{tabular}       \\
\multicolumn{2}{c}{\textit{Memory Hierarchy}}     \\ \hline
L1 I-Cache    & 16 KB, 4-way, Blocking, 2 LatencyCycles
\\
L1 D-Cache    & 16 KB, 4-way, Blocking, 2 LatencyCycles
\\
L2 Cache    & 512 KB, 8-way,  8 MSHRs, 40 LatencyCycles
          \\ \hline
\end{tabular}
\label{table:Setup}
\end{table}

\subsection{Performance Overhead}

\noindent \textbf{Experiment setup.} 
We executed SPECint and Parsec with dual-core verification and compared the performance slowdown of Flexstep with the most widely used hardware scheme Lockstep, and the only available open-source software error-detection scheme Nzdc ~\cite{didehban2016nzdc} which fails to compile on some workloads (\eg, bodytrack, ferret, gcc).
 

 

\noindent \textbf{Results.} The normalised slowdown of running Parsec and SPECint are shown in Fig.~\ref{fig:perf_parsec} and  Fig.~\ref{fig:perf_spec}, respectively. FlexStep incurs a performance slowdown (geomean) of only 1.07\% when running Parsec and 1.24\% when running SPEC. This slowdown is attributed to the extraction of Register Checkpoints and backpressure from Data Buffering resulting from cores undergoing different kernel mode switches and instruction executions. 
In contrast, the slowdown of Nzdc is almost 57.7\% in Parsec and 91.5\% in SPEC, roughly 1.56x and 1.89x slower than FlexStep. Nzdc experiences a significant slowdown due to the need for redundant checking instructions, resulting in substantial performance degradation. While LockStep incurs no performance drop, it requires an additional equivalent area. FlexStep achieves minimal slowdown with reasonable area overheads (Sec.~\ref{area}), highlighting its advantages in both performance and resource efficiency.


We further compare the slowdown of Parsec under different verification modes in Fig.~\ref{diffmode}. Results indicate that the slowdown (geomean) in dual-core mode is 1.07\%, while in triple-core mode, it increases slightly to 1.77\%. This modest increase is attributed to having more checker cores, which exacerbates the execution inconsistency between cores, leading to more frequent backpressure on the main core and affecting its normal execution.
Overall, the performance overhead in both modes remain minimal, demonstrating that FlexStep's support for different verification modes is highly performance-efficient.

\subsection{Percentage of Schedulable Task Sets}
\noindent \textbf{Experiment setup.} 
To assess the schedulability improvements enabled by FlexStep, we conducted numerical simulations using randomly generated task sets across various system configurations. Task sets were generated following the UUnifast algorithm~\cite{Bini_2005}. We compared the percentage of schedulable task sets under three error detection schemes: LockStep, HMR, and FlexStep, where partitioned EDF is used for allocation and scheduling.
For LockStep, tasks with different reliability levels were partitioned into separate queues and ordered by descending utilisation. Verification tasks were allocated first, minimising the number of checker cores required by allocating a new group of main and checker cores only when the current group was fully utilised. Non-verification tasks were allocated last across all available cores.
For HMR, verification tasks were also prioritised during allocation. Non-verification tasks were subsequently assigned, first filling cores without verification tasks and then allocating them to the core with the lowest utilisation.
For FlexStep, the task partitioning is described in Sec.~\ref{theory}.

\noindent \textbf{Results.} 
The schedulability results in Fig.~\ref{sr} demonstrate that FlexStep consistently outperforms LockStep and HMR, especially at higher utilisation levels. As utilisation increases, FlexStep and HMR exhibit a more gradual performance decline, whereas LockStep experiences a sharp drop. This difference arises from LockStep's rigid verification approach, which consumes substantial resources.
HMR encounters few deadline misses even under moderate utilisation, mainly due to the fact that non-verification tasks cannot preempt verification tasks allocated on the same core.
FlexStep’s flexibility allows for more efficient resource allocation, mitigating these issues.

Comparing Figs.~\ref{fig:160-8-10-10}, \ref{fig:160-8-20-20}, and \ref{fig:160-8-40-40}, it is evident that FlexStep achieves greater performance improvements when there are fewer verification tasks, as more non-verification tasks are required to share cores with verification tasks. 
As FlexStep allows flexible switching between verification and non-verification tasks, without requiring binding between main and checker cores, it enhances resource utilisation and improves system schedulability.
The improvement persists with an increased number of cores or higher utilisations of individual tasks, as shown in Figs.~\ref{fig:160-16-20-20} and \ref{fig:80-8-20-20}.
Comparing Figs.~\ref{fig:160-8-20-20} and \ref{fig:160-8-0-40}, the addition of triple-check tasks significantly increases resource demand, leading to varying degrees of performance degradation across all three methods. However, LockStep and HMR exhibit a sharper decline in performance compared to FlexStep, as their core binding between main and checker cores limits flexibility in scheduling. In contrast, FlexStep enables more flexible verification and scheduling, optimising resource utilisation to meet deadlines, even in more demanding scenarios.

\begin{figure}[t]
\begin{center}
\includegraphics[width=0.9\linewidth]{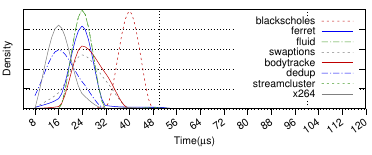}
\caption{Probability distribution of error detection latency of Parsec.}
\label{density}
\end{center}
\end{figure}

\begin{figure}[t]
\begin{center}
\subfigure[Average Power.]{
\hspace{-15pt}
\includegraphics[width=0.46\linewidth]{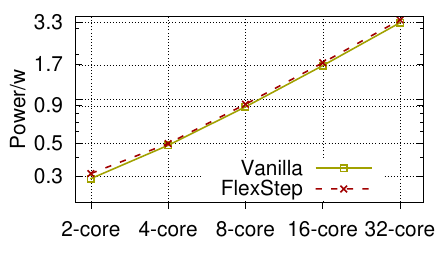}
\label{fig:scal_power}
}
\hspace{-4pt}
\subfigure[Area.]{
\includegraphics[width=0.46\linewidth]{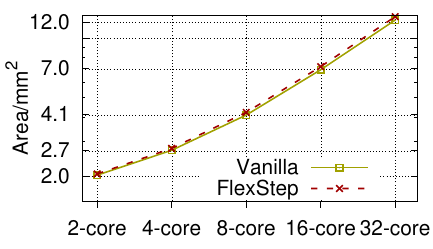}
\label{fig:scal_area}
}
\caption{The average power and area overheads on the SoC with different core numbers for Vanilla and FlexStep, including L1\$ and L2\$.}
\label{scalability}
\end{center}
\end{figure}

\subsection{Error Detection Latency}
\noindent \textbf{Experiment setup.} 
To evaluate detection latency, we injected errors in the forwarded data from the main core, \eg, memory access data of MAL and architectural register data of ASS, simulating the hardware faults without disrupting the main core's normal execution. For each workload, 5,000 - 10,000 faults were randomly generated, resulting in over 100,000 sample points in total to ensure the validity of results.

\noindent\textbf{Results.} Fig.~\ref{density} shows the density distribution of detection latency across different workloads. For most workloads, the majority of detection latencies are concentrated around 20 $\mu$s, while the maximum latency seen in blackscholes is 2 to 3 times higher, reaching up to 50 $\mu$s. Overall, FlexStep maintains an actual error detection latency of no more than 50$\mu$s in most cases, which is sufficient to cover over 99.9\% of hardware faults and guarantee the system reliability.

\subsection{Scalability}

\noindent \textbf{Experiment setup.} 
To evaluate scalability, we increased the number of cores for both FlexStep and Vanilla (based on the original unmodified microarchitecture). We assessed their area and average power consumption using Design Compiler (v2019.12) for RTL synthesis and PrimeTime PX (v2019.12) for post-simulation analysis.


\noindent \textbf{Results.} Fig.~\ref{scalability} presents a comparison of average power (Fig.~\ref{fig:scal_power}) and area (Fig.~\ref{fig:scal_area}) for Vanilla and FlexStep across varying core counts. Results indicate that the increase in average power and area for FlexStep relative to Vanilla remains nearly linear, rather than exponential, as the SoC scales from dual-core to 32-core. This demonstrates that FlexStep offers strong scalability in multi-core and many-core systems.

\subsection{Hardware Overheads}
\label{area}
\noindent \textbf{Experiment setup.} 
We used the same setup as in scalability evaluation to evaluate the hardware overheads of FlexStep compared to Vanilla.

\noindent \textbf{Results.}
With the reuse of homogeneous checker cores, FlexStep introduces minimal hardware overhead. Compared to Vanilla, FlexStep incurs the storage overhead per core of only 1614 bytes: 8 bytes for CPC, 518 bytes for ASS, and 1088 bytes for DBC. For a 4-core SoC shown in Tab.~\ref{table:hardware}, FlexStep occupies 2.77 mm$^2$ of area and consumes 0.499w of power, representing just 2.21\% and 2.89\% overhead, respectively, compared to Vanilla. This validates that the hardware overhead for implementing FlexStep is well within reasonable limits.

\begin{table}[t]
\scriptsize
\tikzset{
    purplebox/.style = {draw =gray, fill = gray!15, rounded corners = 4pt}
}
\caption{Average power \& area of Vanilla and FlexStep (4 cores).}
\centering
\label{table:hardware}
\renewcommand{\arraystretch}{0.9}
\resizebox{0.8\columnwidth}{!}{%
\begin{tabular}{c c c c c}
\hline
\toprule
\multicolumn{1}{c}{} & \multicolumn{1}{c}{\textbf{Vanilla}} & \multicolumn{1}{c}{\textbf{FlexStep}} & \multicolumn{2}{c}{} \\ 
\hline
\toprule
\multicolumn{1}{c|}{Core}    & \multicolumn{1}{c}{Rocket} & \multicolumn{1}{c|}{Rocket} & \multicolumn{1}{c|}{}& \multicolumn{1}{c}{$\times$}  \\

\multicolumn{1}{c|}{Tech. (nm)}    & \multicolumn{1}{c}{28} & \multicolumn{1}{c|}{28} & \multicolumn{1}{c|}{} & \multicolumn{1}{c}{$\times$} \\

\multicolumn{1}{c|}{Power (w)}     & \multicolumn{1}{c}{0.485}      & \multicolumn{1}{c|}{0.499} & \multicolumn{1}{c|}{} & \multicolumn{1}{c}{2.89\%}       \\

\multicolumn{1}{c|}{Area (mm$^2$)}     & \multicolumn{1}{c}{2.71}      & \multicolumn{1}{c|}{2.77} &\multicolumn{1}{c|}{\multirow{-4.5}{*}{\rotatebox[origin=c]{270}{\textbf{Overhead}}}} & \multicolumn{1}{c}{2.21\%}       \\

\hline
\toprule
\end{tabular}
}
\end{table}

\section{conclusion}

In this work, we developed FlexStep, a homogeneous-core error-detection framework. Using a hardware-software co-design approach, we implemented FlexStep in a multi-core system by adding lightweight modifications to existing microarchitecture of Rocket and integrating simple OS-level scheduling codes. Evaluated on an FPGA prototype, FlexStep demonstrates microsecond-level error detection capabilities with low performance slowdown and hardware overheads, high scalability, and dynamic verification mode switching, making it an efficient solution for practical applications. In future works, FlexStep enables flexible task scheduling and unlocks significant potential for developing new scheduling algorithms and analyses to enhance system efficiency, improve real-time responsiveness, and adapt to dynamic reliability requirements in modern safety-critical applications.

\section{Acknowledgement}
We would like to thank the editors and anonymous reviewers for their helpful feedback. This work is supported by National Key Research and Development Program (Grant No.2024YFB4405600), National Key Research and Development Program of China (Grant No.2022ZD0118902), U.S. National Science Foundation (Grant No.CNS-2340171), the National Natural Science Foundation of China (Grant No. 62472086), National Natural Science Foundation of China (Grant Nos.92264203, 62204036), the Basic Research Program of Jiangsu (Grants No. BK20243042), Key Research and Development Program of Jiangsu Province(Grant No.BE2023020-1), and the Start-up Research Fund of Southeast University (Grant No. RF1028624005).

\bibliographystyle{IEEEtran}
\bibliography{ref}

\begin{thebibliography}{10}
\providecommand{\url}[1]{#1}
\csname url@samestyle\endcsname
\providecommand{\newblock}{\relax}
\providecommand{\bibinfo}[2]{#2}
\providecommand{\BIBentrySTDinterwordspacing}{\spaceskip=0pt\relax}
\providecommand{\BIBentryALTinterwordstretchfactor}{4}
\providecommand{\BIBentryALTinterwordspacing}{\spaceskip=\fontdimen2\font plus
\BIBentryALTinterwordstretchfactor\fontdimen3\font minus \fontdimen4\font\relax}
\providecommand{\BIBforeignlanguage}[2]{{%
\expandafter\ifx\csname l@#1\endcsname\relax
\typeout{** WARNING: IEEEtran.bst: No hyphenation pattern has been}%
\typeout{** loaded for the language `#1'. Using the pattern for}%
\typeout{** the default language instead.}%
\else
\language=\csname l@#1\endcsname
\fi
#2}}
\providecommand{\BIBdecl}{\relax}
\BIBdecl

\bibitem{rausand2014reliability}
M.~Rausand, \emph{Reliability of safety-critical systems: theory and applications}.\hskip 1em plus 0.5em minus 0.4em\relax John Wiley \& Sons, 2014.

\bibitem{kumar2018reliability}
V.~Kumar, L.~Singh, and A.~K. Tripathi, ``Reliability analysis of safety-critical and control systems: a state-of-the-art review,'' \emph{IET Software}, vol.~12, no.~1, pp. 1--18, 2018.

\bibitem{maurya2020reliability}
A.~Maurya and D.~Kumar, ``Reliability of safety-critical systems: A state-of-the-art review,'' \emph{Quality and Reliability Engineering International}, vol.~36, no.~7, pp. 2547--2568, 2020.

\bibitem{singh2021reliability}
P.~Singh and L.~K. Singh, ``Reliability and safety engineering for safety critical systems: An interview study with industry practitioners,'' \emph{IEEE Transactions on Reliability}, vol.~70, no.~2, pp. 643--653, 2021.

\bibitem{kumar2022reliability}
V.~Kumar, K.~C. Mishra, P.~Singh, A.~N. Hati, M.~R. Mamdikar, L.~K. Singh, and R.~R. Parida, ``Reliability analysis and safety model checking of safety-critical and control systems: A case study of npp control system,'' \emph{Annals of Nuclear Energy}, vol. 166, p. 108812, 2022.

\bibitem{huang2014scheduling}
P.~Huang, H.~Yang, and L.~Thiele, ``On the scheduling of fault-tolerant mixed-criticality systems,'' in \emph{Proceedings of the 51st annual design automation conference}, 2014, pp. 1--6.

\bibitem{ali2019rt-scheduling}
W.~Ali and H.~Yun, ``Rt-gang: Real-time gang scheduling framework for safety-critical systems,'' in \emph{2019 IEEE Real-Time and Embedded Technology and Applications Symposium (RTAS)}.\hskip 1em plus 0.5em minus 0.4em\relax IEEE, 2019, pp. 143--155.

\bibitem{kumar2020dynamic-scheduling}
V.~P. Kumar and A.~S. Pillai, ``Dynamic scheduling algorithm for a utomotive safety critical systems,'' in \emph{2020 Fourth International Conference on Computing Methodologies and Communication (ICCMC)}.\hskip 1em plus 0.5em minus 0.4em\relax IEEE, 2020, pp. 815--820.

\bibitem{melani2017static-scheduling}
A.~Melani, M.~A. Serrano, M.~Bertogna, I.~Cerutti, E.~Quinones, and G.~Buttazzo, ``A static scheduling approach to enable safety-critical openmp applications,'' in \emph{2017 22nd Asia and South Pacific Design Automation Conference (ASP-DAC)}.\hskip 1em plus 0.5em minus 0.4em\relax IEEE, 2017, pp. 659--665.

\bibitem{calvaresi2018local-scheduling}
D.~Calvaresi, M.~Marinoni, L.~Lustrissimini, K.~Appoggetti, P.~Sernani, A.~F. Dragoni, M.~Schumacher, and G.~Buttazzo, ``Local scheduling in multi-agent systems: getting ready for safety-critical scenarios,'' in \emph{Multi-Agent Systems and Agreement Technologies: 15th European Conference, EUMAS 2017, and 5th International Conference, AT 2017, Evry, France, December 14-15, 2017, Revised Selected Papers 15}.\hskip 1em plus 0.5em minus 0.4em\relax Springer, 2018, pp. 96--111.

\bibitem{dual-lock}
A.~Hanafi, M.~Karim, and A.~E. Hammami, ``Dual-lockstep microblaze-based embedded system for error detection and recovery with reconfiguration technique,'' in \emph{2015 Third World Conference on Complex Systems (WCCS)}, 2015, pp. 1--6.

\bibitem{arm-triple-lock}
X.~Iturbe, B.~Venu, E.~Ozer, and S.~Das, ``A triple core lock-step (tcls) arm® cortex®-r5 processor for safety-critical and ultra-reliable applications,'' in \emph{2016 46th Annual IEEE/IFIP International Conference on Dependable Systems and Networks Workshop (DSN-W)}, 2016, pp. 246--249.

\bibitem{iturbe2019arm}
X.~Iturbe, B.~Venu, E.~Ozer, J.-L. Poupat, G.~Gimenez, and H.-U. Zurek, ``The arm triple core lock-step (tcls) processor,'' \emph{ACM Transactions on Computer Systems (TOCS)}, vol.~36, no.~3, pp. 1--30, 2019.

\bibitem{werdmuller2015addressing}
N.~Werdmuller, ``Addressing functional safety applications with arm cortex-r5,'' 2015.

\bibitem{iturbe2018addressing-lock}
X.~Iturbe, B.~Venu, J.~Jagst, E.~Ozer, P.~Harrod, C.~Turner, and J.~Penton, ``Addressing functional safety challenges in autonomous vehicles with the arm tcl s architecture,'' \emph{IEEE Design \& Test}, vol.~35, no.~3, pp. 7--14, 2018.

\bibitem{Cortex-R5}
{ARM}, \emph{Cortex-R5 and Cortex-R5F Technical Reference Manual}, \url{https://developer.arm.com}, 2024.

\bibitem{hukerikar2018redthreads-soft}
S.~Hukerikar, K.~Teranishi, P.~C. Diniz, and R.~F. Lucas, ``Redthreads: An interface for application-level fault detection/correction through adaptive redundant multithreading,'' \emph{International Journal of Parallel Programming}, vol.~46, pp. 225--251, 2018.

\bibitem{bernardi2015development-soft}
P.~Bernardi, R.~Cantoro, S.~De~Luca, E.~S{\'a}nchez, and A.~Sansonetti, ``Development flow for on-line core self-test of automotive microcontrollers,'' \emph{IEEE Transactions on Computers}, vol.~65, no.~3, pp. 744--754, 2015.

\bibitem{sofycomputation-soft}
D.~S. Khudia and S.~Mahlke, ``Harnessing soft computations for low-budget fault tolerance,'' in \emph{2014 47th Annual IEEE/ACM International Symposium on Microarchitecture}, 2014, pp. 319--330.

\bibitem{thomas2013error-soft}
A.~Thomas and K.~Pattabiraman, ``Error detector placement for soft computation,'' in \emph{2013 43rd Annual IEEE/IFIP International Conference on Dependable Systems and Networks (DSN)}.\hskip 1em plus 0.5em minus 0.4em\relax IEEE, 2013, pp. 1--12.

\bibitem{ismael2021scheduling-SA}
G.~A. Ismael, A.~A. Salih, A.~AL-Zebari, N.~Omar, K.~J. Merceedi, A.~J. Ahmed, N.~O. Salim, S.~S. Hasan, S.~F. Kak, I.~M. Ibrahim \emph{et~al.}, ``Scheduling algorithms implementation for real time operating systems: A review,'' \emph{Asian Journal of Research in Computer Science}, vol.~11, no.~4, pp. 35--51, 2021.

\bibitem{agrawal2021cpu-SA}
P.~Agrawal, A.~K. Gupta, and P.~Mathur, ``Cpu scheduling in operating system: a review,'' in \emph{Proceedings of the Second International Conference on Information Management and Machine Intelligence: ICIMMI 2020}.\hskip 1em plus 0.5em minus 0.4em\relax Springer, 2021, pp. 279--289.

\bibitem{hamayun2015optimized-SA}
M.~Hamayun and H.~Khurshid, ``An optimized shortest job first scheduling algorithm for cpu scheduling,'' \emph{J. Appl. Environ. Biol. Sci}, vol.~5, no.~12, pp. 42--46, 2015.

\bibitem{6899155-trend}
F.~Kluge, M.~Gerdes, and T.~Ungerer, ``An operating system for safety-critical applications on manycore processors,'' in \emph{2014 IEEE 17th International Symposium on Object/Component/Service-Oriented Real-Time Distributed Computing}, 2014, pp. 238--245.

\bibitem{el2013across-trend2}
C.~El~Salloum, M.~Elshuber, O.~H{\"o}ftberger, H.~Isakovic, and A.~Wasicek, ``The across mpsoc--a new generation of multi-core processors designed for safety--critical embedded systems,'' \emph{Microprocessors and Microsystems}, vol.~37, no.~8, pp. 1020--1032, 2013.

\bibitem{saidi2015shift-trend3}
S.~Saidi, R.~Ernst, S.~Uhrig, H.~Theiling, and B.~D. de~Dinechin, ``The shift to multicores in real-time and safety-critical systems,'' in \emph{2015 International Conference on Hardware/Software Codesign and System Synthesis (CODES+ ISSS)}.\hskip 1em plus 0.5em minus 0.4em\relax IEEE, 2015, pp. 220--229.

\bibitem{cerrolaza2020multi-trend4}
J.~P. Cerrolaza, R.~Obermaisser, J.~Abella, F.~J. Cazorla, K.~Gr{\"u}ttner, I.~Agirre, H.~Ahmadian, and I.~Allende, ``Multi-core devices for safety-critical systems: A survey,'' \emph{ACM Computing Surveys (CSUR)}, vol.~53, no.~4, pp. 1--38, 2020.

\bibitem{perez2022gpu-trend5}
J.~Perez-Cerrolaza, J.~Abella, L.~Kosmidis, A.~J. Calderon, F.~Cazorla, and J.~L. Flores, ``Gpu devices for safety-critical systems: A survey,'' \emph{ACM Computing Surveys}, vol.~55, no.~7, pp. 1--37, 2022.

\bibitem{durrieu2014predictable-trend6}
G.~Durrieu, M.~Faug{\`e}re, S.~Girbal, D.~G. P{\'e}rez, C.~Pagetti, and W.~Puffitsch, ``Predictable flight management system implementation on a multicore processor,'' in \emph{Embedded Real Time Software (ERTS'14)}, 2014.

\bibitem{rogenmoser2023hybrid}
M.~Rogenmoser, Y.~Tortorella, D.~Rossi, F.~Conti, and L.~Benini, ``Hybrid modular redundancy: Exploring modular redundancy approaches in risc-v multi-core computing clusters for reliable processing in space,'' \emph{ACM Transactions on Cyber-Physical Systems}, 2023.

\bibitem{Cortex-A76AE}
ARM, \emph{Arm Cortex-A76AE Core Technical Reference Manual}, 2011.

\bibitem{kempf2022holistic-splitlock}
F.~Kempf, C.~K{\"u}hbacher, C.~Mellwig, S.~Altmeyer, T.~Ungerer, and J.~Becker, ``A holistic hardware-software approach for fault-aware embedded systems,'' in \emph{2022 25th Euromicro Conference on Digital System Design (DSD)}.\hskip 1em plus 0.5em minus 0.4em\relax IEEE, 2022, pp. 704--711.

\bibitem{kempf2021adaptive-flexible}
F.~Kempf, T.~Hartmann, S.~Baehr, and J.~Becker, ``An adaptive lockstep architecture for mixed-criticality systems,'' in \emph{2021 IEEE Computer Society Annual Symposium on VLSI (ISVLSI)}.\hskip 1em plus 0.5em minus 0.4em\relax IEEE, 2021, pp. 7--12.

\bibitem{jeffery2010flexible}
C.~M. Jeffery and R.~J. Figueiredo, ``A flexible approach to improving system reliability with virtual lockstep,'' \emph{IEEE Transactions on Dependable and Secure Computing}, vol.~9, no.~1, pp. 2--15, 2010.

\bibitem{bas2021safede-flexible}
F.~Bas, S.~Alcaide, R.~Lorenzo, G.~Cabo, G.~Gil, O.~Sala, F.~Mazzocchetti, D.~Trilla, and J.~Abella, ``Safede: a flexible diversity enforcement hardware module for light-lockstepping,'' in \emph{2021 IEEE 27th International Symposium on On-Line Testing and Robust System Design (IOLTS)}.\hskip 1em plus 0.5em minus 0.4em\relax IEEE, 2021, pp. 1--7.

\bibitem{barbirotta2024dynamic-flexible}
M.~Barbirotta, F.~Menichelli, A.~Cheikh, A.~Mastrandrea, M.~Angioli, and M.~Olivieri, ``Dynamic triple modular redundancy in interleaved hardware threads: An alternative solution to lockstep multi-cores for fault-tolerant systems,'' \emph{IEEE Access}, 2024.

\bibitem{asanovic2016rocket}
K.~Asanovic, R.~Avizienis, J.~Bachrach, S.~Beamer, D.~Biancolin, C.~Celio, H.~Cook, D.~Dabbelt, J.~Hauser, A.~Izraelevitz \emph{et~al.}, ``The rocket chip generator,'' \emph{EECS Department, University of California, Berkeley, Tech. Rep. UCB/EECS-2016-17}, vol.~4, pp. 6--2, 2016.

\bibitem{ainsworth2018parallel}
S.~Ainsworth and T.~M. Jones, ``Parallel error detection using heterogeneous cores,'' in \emph{2018 48th Annual IEEE/IFIP International Conference on Dependable Systems and Networks (DSN)}.\hskip 1em plus 0.5em minus 0.4em\relax IEEE, 2018, pp. 338--349.

\bibitem{ainsworth2021paradox}
S.~Ainsworth, L.~Zoubritzky, A.~Mycroft, and T.~M. Jones, ``Paradox: Eliminating voltage margins via heterogeneous fault tolerance,'' in \emph{2021 IEEE International Symposium on High-Performance Computer Architecture (HPCA)}.\hskip 1em plus 0.5em minus 0.4em\relax IEEE, 2021, pp. 520--532.

\bibitem{ainsworth2019paramedic}
S.~Ainsworth and T.~M. Jones, ``Paramedic: Heterogeneous parallel error correction,'' in \emph{2019 49th Annual IEEE/IFIP International Conference on Dependable Systems and Networks (DSN)}.\hskip 1em plus 0.5em minus 0.4em\relax IEEE, 2019, pp. 201--213.

\bibitem{chipyard}
A.~Amid, D.~Biancolin, A.~Gonzalez \emph{et~al.}, ``Chipyard: Integrated design, simulation, and implementation framework for custom socs,'' \emph{IEEE Micro}, vol.~40, pp. 10--21, 2020.

\bibitem{TSMC_28}
``{28nm PDKs},'' \url{https://www.tsmc.com/english/dedicatedFoundry/technology/logic/l_28nm}.

\bibitem{karandikar2018firesim}
S.~Karandikar, H.~Mao, D.~Kim, D.~Biancolin, A.~Amid, D.~Lee, N.~Pemberton, E.~Amaro, C.~Schmidt, A.~Chopra \emph{et~al.}, ``Firesim: Fpga-accelerated cycle-exact scale-out system simulation in the public cloud,'' in \emph{2018 ACM/IEEE 45th Annual International Symposium on Computer Architecture (ISCA)}.\hskip 1em plus 0.5em minus 0.4em\relax IEEE, 2018, pp. 29--42.

\bibitem{henning2006spec}
J.~L. Henning, ``Spec cpu2006 benchmark descriptions,'' \emph{ACM SIGARCH Computer Architecture News}, vol.~34, no.~4, pp. 1--17, 2006.

\bibitem{bienia2008parsec}
C.~Bienia, S.~Kumar, J.~P. Singh, and K.~Li, ``The parsec benchmark suite: Characterization and architectural implications,'' in \emph{Proceedings of the 17th international conference on Parallel architectures and compilation techniques}, 2008, pp. 72--81.

\bibitem{didehban2016nzdc}
M.~Didehban and A.~Shrivastava, ``nzdc: A compiler technique for near zero silent data corruption,'' in \emph{Proceedings of the 53rd Annual Design Automation Conference}, 2016, pp. 1--6.

\bibitem{Bini_2005}
E.~Bini and G.~C. Buttazzo, ``Measuring the performance of schedulability tests,'' \emph{Real-Time Systems (Real-Time Syst.)}, vol.~30, no. 1-2, pp. 129--154, 2005.

\end{thebibliography}
\end{document}